\documentclass[11pt]{article}
\baselineskip
\usepackage{amssymb}
\usepackage{graphicx}
\usepackage{latexsym,cite}
\usepackage{epsfig}
\usepackage{amsmath, bbm, color}

\newcommand{\N}{\mathbb{N}}
\newcommand{\Z}{\mathbb{Z}}

\newcommand{\SU}{\mathrm{SU}}

\newcommand{\re}{{\rm{Re}}}

\newcommand{\dd}{{\rm{d}}}
\newcommand{\Tr}{{\rm Tr\,}}
\newcommand{\Seff}{S_{\mbox{\tiny{eff}}}}
\newcommand{\Sng}{S_{\mbox{\tiny{NG}}}}
\newcommand{\Eclosed}{E^{\mbox{\tiny{c}}}}
\newcommand{\Eopen}{E^{\mbox{\tiny{o}}}}

\newcommand{\nconf}{n_{\mbox{\tiny{conf}}}}
\newcommand{\rmin}{r_{\mbox{\tiny{min}}}}
\newcommand{\betagauge}{\beta_{\tiny\mbox{gauge}}}
\newcommand{\betaspin}{\beta_{\tiny\mbox{spin}}}
\newcommand{\Se}{S^{\tiny\mbox{E}}}
\newcommand{\Sel}{S^{\tiny\mbox{E}}_{\tiny\mbox{L}}}
\newcommand{\Glat}{G_{\tiny\mbox{L}}}
\newcommand{\Zlat}{Z_{\tiny\mbox{L}}}
\newcommand{\redchisq}{\chi^2_{\tiny\mbox{red}}}
\newcommand{\eq}{\begin{equation}}
\newcommand{\en}{\end{equation}}
\newcommand{\eqar}{\begin{eqnarray}}
\newcommand{\enar}{\end{eqnarray}}

\begin{document}

\begin{titlepage}
\begin{flushright} 
DFTT 05/11\\ 
HIP-2011-03/TH\\ 
\end{flushright} 
\vskip0.5cm 
\begin{center}
{\Large\bf
Universal signatures of the effective string in finite temperature lattice gauge theories
}
\end{center}
\vskip1.3cm
\centerline{Michele~Caselle$^{a}$, Alessandra~Feo$^{a}$, Marco~Panero$^{b}$ and Roberto~Pellegrini$^{a}$}
\vskip1.5cm
\centerline{\sl  $^a$ Dipartimento di Fisica Teorica dell'Universit\`a di Torino and INFN, Sezione di Torino,}    
 \centerline{\sl Via P.~Giuria 1, I-10125 Torino, Italy}
\vskip0.5cm
\centerline{\sl  $^b$ Department of Physics and Helsinki Institute of Physics, University of Helsinki,}    
 \centerline{\sl FIN-00014 Helsinki, Finland}
\vskip0.5cm
\begin{center}
{\sl  E-mail:} \hskip 5mm \texttt{caselle@to.infn.it, feo@to.infn.it, marco.panero@helsinki.fi, pellegri@to.infn.it}
\end{center}
\vskip1.0cm
\begin{abstract}

We study the behavior of the interquark potential in  lattice gauge theories at high temperature, but still in the 
confining phase, and propose a new observable which could play in this regime 
the same r\^ole played by the L\"uscher term
 in the  low temperature limit.  
 This quantity is related to the exponent of the power prefactor in the effective string partition function or, equivalently, 
to the coefficient of the logarithmic correction in the interquark potential and,
 as for the usual  L\"uscher term, its value does not depend on the particular gauge group under consideration or on the form of the effective string action used
 to model the flux tube. In this respect it can be considered as a universal signature of the effective string behavior of the flux tube.
As a test of our proposal we studied this quantity with a set of 
high-precision numerical simulations in the (2+1) dimensional $\SU(2)$, $\SU(3)$ and $\SU(4)$ Yang-Mills theories and in the $\Z_2$ gauge model, always 
finding a perfect agreement with the predicted values.

\end{abstract}
\vspace*{0.2cm}

\end{titlepage}

\section{Introduction and motivation} 
\label{introsect} 

It is by now rather well understood that the flux tube joining a static quark-antiquark pair in 
the confining regime of a generic lattice gauge theory (LGT) 
can be described by an effective string theory~\cite{Luescherterm}.
The most relevant consequence of this description is the presence of a term in the interquark potential proportional to $1/r$ (where $r$ is the interquark distance) 
with a fixed coefficient (the well known ``L\"uscher term''~\cite{Luescherterm}) which can be predicted from the theory and is not 
affected by corrections due to higher order terms in the effective string action. 
In this sense the L\"uscher term
can be considered as a universal signature of the effective string behavior of the flux tube and in the past years it was the first goal of any numerical study on the
subject. More recently it has been realized that this term is only the first of a series of universal corrections in the perturbative expansion of the
 effective string action in powers of $\sigma r^2$~\cite{lw04,Aharony}  opening the way to more refined numerical test of the effective string picture
 (for an updated review of lattice results, see~\cite{Teper:2009uf}).

While the first numerical tests, more than twenty years ago, were performed using Wilson loops~\cite{wloop}, in these last years it has become more frequent to choose a finite
temperature setting and look at the interquark potential using Polyakov loop correlators (or torelons, depending to the choice of the space and time directions on the
lattice). 
These finite temperature tests can be performed in two very different regimes. One of them is the ``open string channel'', which in the LGT language is the low temperature regime: it 
corresponds to lattices with a very long size in the ``inverse temperature'' direction and hence very long Polyakov loops.  In this regime the dominant 
effective string
correction to the potential is again the  L\"uscher term and higher order (possibly universal) string corrections appear as contributions with 
higher powers in $1/r$ in the
interquark potential. The other possible choice is the ``closed string channel'', which in the LGT language corresponds to the high temperature (but still in the confining
region) regime. In this case the size of the lattice in the compactified direction is just above the deconfinement length and
 the Polyakov loops are much shorter than the typical interquark distance. In this regime the dominant effective string correction is linear in $r$ and has the effect of
 decreasing the string tension which becomes temperature dependent. As the temperature increases the lattice size in the compactified direction
 becomes shorter and shorter and accordingly the string tension decreases and at
 the end it vanishes leading to the deconfinement transition. 
 From a physical point of view, this high $T$ regime is particularly interesting 
 since in this limit the behavior of the flux tube and of its quantum fluctuations could strongly influence the transition from hadrons to deconfined quarks as $T_c$ is approached.  
 The fact that the L\"uscher term gets mixed with higher order corrections makes it difficult to use it as a universal signature of the presence of an effective string.
 Indeed in this regime one usually has to compare the numerical results with the expectation of the whole effective string action which is not universal anymore\footnote{Indeed in this limit the typical question is how accurate are the results from the Nambu-Goto model, which is the simplest possible effective string theory
 and in several cases turns out to fit the numerical data very well~\cite{Teper:2009uf}.}.
 
 Also in this regime, it would be interesting to have a universal signature of the effective string which could play the same r\^ole played by the L\"uscher term
 in the  low temperature (``open string'') limit. 
  As we shall show in detail below, the natural candidate for such a r\^ole is the coefficient of the logarithmic
 correction in the interquark potential (or, equivalently, the exponent of the powerlike prefactor in front of the effective string partition function).
 This term, which grows logarithmically with $r$, is not affected by higher order contributions of the effective
 string action and only depends on the nature of the conformal field theory (CFT) to which the effective string model flows in the infrared limit. 
 In particular,
 the presence of this term is a signature of the free bosonic nature of the quantum fluctuations of the flux tube in this limit 
 i.e. the fact that they can be described  by the $c=1$ CFT of a free \emph{uncompactified} 
 bosonic degree of freedom. In this respect it can be considered as a high temperature analogue of the L\"uscher term and,
 exactly as the L\"uscher term, it is universal  and has a simple linear dependence on the number
 of transverse dimensions. 

 The main  goal of the present paper is to study this observable and test its universality in 
 a set of LGTs both with continuous and discrete gauge groups. In this respect
 it is important to note that a projection of the Polyakov loops onto their 
 zero transverse-momentum component (which is normally performed in this type of lattice computations~\cite{Teper:2009uf}) would wash 
 away such a term. Therefore in our computation we did not perform a projection onto the zero transverse-momentum 
 component, and the correlators we investigated also involve higher momenta. This demands a special computational 
 effort, and a combination of  state-of-the-art simulation and error-reduction algorithms, with large statistics. 
 
 As we shall see, in all the cases we studied 
 (which are certainly described by different effective string actions, since their critical behavior in the
 vicinity of the deconfinement point is different), the coefficient of this logarithmic term in the potential
  is always the same and it agrees with the predicted value. This makes this observable a robust signature of the effective string behavior 
  in the closed string channel and a good candidate to play the r\^ole of the L\"uscher term in the high $T$ regime of confining LGTs.

This paper is organized as follows. In section~\ref{effective_string_section}, 
we remind the construction and the most important properties of the bosonic effective string model, discussing, in particular, the Nambu-Goto action. 
Section~\ref{2+1_gauge_theories_section} recalls the basic features of Yang-Mills theories in $2+1$ dimensions and introduces the setup of our 
lattice simulations, whose results are reported in section~\ref{results_section}. Section~\ref{conclusions_section} presents a discussion, 
and some concluding remarks.

\section{Effective string theory}
\label{effective_string_section}

In the confined phase of Yang-Mills theories, the zero-temperature ground state interquark potential $V_0(r)$ for a pair of fundamental, static color sources $Q\bar{Q}$ at asymptotically large distances $r$ is linearly rising with $r$: this can be interpreted as evidence for the formation of a stable ``tube'' of chromoelectric flux lines between the two charges, with an energy proportional to its length. In the infrared limit, such tube can be interpreted as a string-like object, with negligible intrinsic width and no transverse structure. If there is one such string in the confining Yang-Mills vacuum, then the translational and rotational symmetries of spacetime get spontaneously broken\footnote{The breaking is actually explicit, if the sources are not located at infinity.} down to the subgroup of translations parallel to and rotations around the string world sheet. This leads to the existence of $D-2$ massless Nambu-Goldstone bosons (the transverse fluctuation modes of the string), which are expected to be the relevant degrees of freedom of an effective theory, valid at energy scales well below the intrinsic mass gap of the confining theory. 

It is important to emphasize that this effective string picture is expected to provide a \emph{universal} low-energy description, i.e., one independent of the underlying gauge theory (as long as it is confining) and of its microscopic dynamics details. 

For an open string of finite length $r$, the quantization of the transverse vibration modes (treated as non-interacting) leads to a Casimir effect, which manifests itself as a $1/r$ term\footnote{Note that such a term has nothing to do with a short-distance Coulomb term.} in the ground-state interquark potential~\cite{Luescherterm}:
\eq
\label{groundstatepotential}
V_0(r) \sim \sigma_0 r - \frac{\pi (D-2)}{24 r},
\en
to a tower of evenly spaced excited levels:
\eq
\label{excitedlevels} 
E_n(r) \sim V_0(r) + \frac{\pi}{r}n, \;\;\; n \in \N,
\en
and to a logarithmic broadening of the mean square width $w^2$ of the fluctuations in the midpoint of the string, as a function of the string length:
\eq
\label{width}
w^2 (r) \sim \frac{D-2}{2 \pi \sigma_0} \ln \left( r/r_0 \right)
\en
(where $r_0$ denotes a fixed length scale).

In order to describe the complete partition function for the sector of the gauge theory containing a pair of static color sources, one needs to generalize these predictions by including string interactions, which are encoded in the complete effective string action $\Seff$. During its time-like evolution in a Euclidean setup---possibly at a non-negligible temperature $T$---, each static source traces out a straight worldline which is a Polyakov loop winding around the compact time direction. The two-point Polyakov loop correlation function $G(r,T)$ is then written as the string partition function:
\eq
\label{Polyakov_correlator}
G(r,T) = \langle \mathcal{P}^* \left(\vec{r}\right) \mathcal{P}(\vec{0}) \rangle = \int \mathcal{D} h  e^{-\Seff[h]},
\en
where the functional integral of the transverse fluctuations of the string $h$ (which are the degrees of freedom in the \emph{physical gauge}) is to be performed over connected world-sheet configurations having the Polyakov loops as their boundary.

The functional form of $\Seff$ is not known \emph{a priori}, but it can be expanded as a series of derivatives of $h$, including both bulk and boundary terms. At each given
order in the number of derivatives and fields, the various terms can be constrained by Poincar\'e symmetry and open-closed string duality~\cite{Aharony} (see also \cite{Dass}). Somewhat unexpectedly, it turns out that the first few orders in this series exactly reproduce the expansion of the Nambu-Goto action~\cite{NambuGoto}, 
which is just proportional to the area of the string world-sheet:
\eq
\label{Nambu_Goto_action}
\Sng=\sigma_0 \int d^2 \mathcal{A}.
\en
The effective string expansion shows deviations from eq.~(\ref{Nambu_Goto_action}) only at high orders~\cite{Aharony, Brandt:2010bw}, which are at the limit or beyond the resolution of current lattice calculations. In particular, for the open string channel in $D=2+1$ dimensions, the leading-order corrections (which are induced by boundary terms appearing in the effective action) are predicted to be proportional to $r^{-4}$, and have been recently investigated in high-precision simulations~\cite{Brandt:2010bw}. For the closed string channel, the leading-order corrections appear at an even higher order, and are proportional to $r^{-7}$. This justifies considering the Nambu-Goto action as a very accurate approximation of the actual effective string action, within the precision of present numerical simulation results.

Thanks to its particularly simple form, the energy levels and properties of the Nambu-Goto action can be easily derived analytically; in particular, the open string spectrum reads~\cite{Arvis:1983fp}:
\eq
\label{Arvis_spectrum}
\Eopen_n=\sigma_0 r \sqrt{ 1+ \frac{2 \pi}{\sigma_0 r^2} \left( n -\frac{D-2}{24} \right)}, \;\;\; n \in \N.
\en
It is well-known that the Nambu-Goto string is generically affected by an anomaly, leading to breakdown of rotational invariance (except for the critical number of spacetime dimensions $D=26$~\cite{Goddard:1973qh}), and that, being non-polynomial, it is non-renormalizable. However, these issues are not of concern to us here, given that in the present context one just regards eq.~(\ref{Nambu_Goto_action}) as (an approximation to) the action for an effective theory valid at low energies only and that the effects of the anomaly in non-critical dimensions are suppressed in the long-string limit~\cite{Olesen:1985pv}.

Approximating $\Seff$ by $\Sng$ on the r.h.s. of eq.~(\ref{Polyakov_correlator}), the Polyakov loop correlation function in $D=2+1$ dimensions at a finite 
temperature $T$ can be expressed as a series of modified Bessel functions. 
This result was obtained 
a few years ago by L\"uscher and Weisz~\cite{lw04} using a duality transformation 
and then derived in the covariant formalism in~\cite{Billo:2005iv}. In $d=2+1$ dimensions one finds a tower of $K_0$ Bessel functions:  
\eq
\label{NG_Polyakov_correlator}
G(r,T) =\sum_{n=0}^{\infty} c_n K_{0}\left( r \Eclosed_n \right) 
\en 
with arguments involving the Nambu-Goto energy levels for a closed string: 
\eq 
\label{closed_string_NG_spectrum}
\Eclosed_n= \frac{\sigma_0}{T}
\sqrt{1+ \frac{8\pi T^2}{\sigma_0} \left( n -\frac{1}{24} \right)}, \;\;\; n \in \N.
\en 
In particular, in the large-distance regime $rT \gg 1$, eq.~(\ref{NG_Polyakov_correlator}) is dominated by the term depending on the lowest energy level:
\eq 
\label{G_dominated_by_K0}
G(r,T) \simeq K_0 \left( r \Eclosed_0 \right).
\en 
Following ref.~\cite{Olesen:1985ej}, one can also define a temperature-dependent string tension $\sigma=\sigma(T)$ as:
\eq 
\label{T-dependent_sigma}
\sigma = \Eclosed_0 T = \sigma_0 \sqrt{ 1 - \frac{\pi T^2}{3\sigma_0} },
\en 
which is predicted to vanish at a critical temperature $T^{\mbox{\tiny{NG}}}_c=\sqrt{3\sigma_0/\pi}$.

Given that the two-point Polyakov loop correlation function $G(r,T)$ can be related to the finite-temperature heavy-quark potential $V(r,T)$ via:
\eq
\label{potential_definition}
V(r,T) = - T \ln G(r,T),
\en
and that the modified Bessel function $K_n(z)$ has the following expansion:
\eq
\label{K0_expansion}
K_n(z) = \sqrt{\frac{\pi}{2z}} e^{-z} \left[ 1 + \frac{4n^2-1}{8z} + \frac{16n^4-40n^2+9}{128z^2}+ \mathcal{O}(z^{-3})  \right]
\en
for large values of its argument $z$, one gets:
\eq
\label{potential_approximate_form}
V(r,T) \sim rT \Eclosed_0 + \frac{T}{2} \ln \left( r \Eclosed_0 \right) + \frac{T}{8 r \Eclosed_0},
\en
where we dropped an irrelevant additive constant, and neglected terms which are suppressed by higher powers of $(rT)^{-1}$. 

At this point, a remark is in order: the corrections to the effective string action can be written in a series of powers of $(\sigma r^2)^{-1}$, where $\sigma$ is the string tension at the temperature $T$, and vanishes in the $T \to T_c^-$ limit, for a second-order phase transition. This imposes a further constraint on the validity range of our expansion, and, in particular, implies that the model cannot be expected to hold all the way up to $T_c$. In our analysis, we shall include only data corresponding to $\sigma r^2 > 1.5$, which---based on previous studies---appears to be a reasonable criterion. 

Within the $r$ and $T$ ranges satisfying these conditions for the validity of the expansion, eq.~(\ref{potential_approximate_form}) predicts thus a characteristic, logarithmic contribution to the confining heavy-quark potential. It is important to stress that, for the values of $r$ and $T$ usually
studied in lattice simulations, this logarithmic term is much larger than the statistical errors. Moreover, as mentioned in the introduction, 
it is universal in the sense that any effective string model which has a free bosonic CFT as its large-$r$ limit
must contain this contribution which can thus be considered as a distinguishing feature of this class of effective string models.
At the same time it is important to stress that such a term would be washed away, 
if one---as it is often done in lattice computations---were to project the Polyakov lines to their zero transverse momentum components. From the point of view of the numerical lattice computation, this makes the observation of the effect more challenging, 
and requires the use of error-reduction techniques which are discussed in the next section. In the remaining part of this paper we shall mainly focus on this contribution.

\section{Confining gauge theories in $2+1$ dimensions and their lattice regularization} 
\label{2+1_gauge_theories_section}

In this section, we first introduce $\SU(N)$ Yang-Mills theories in $2+1$ dimensions in subsection~\ref{2+1_Yang_Mills_subsection}, then we discuss their lattice regularization in subsection~\ref{SUN_lattice_regularization_subsection}, as well as the $\Z_2$ lattice gauge theory (subsection~\ref{Z2_subsection}). Finally, we define the observables evaluated in our lattice simulations in subsection~\ref{observables_subsection}.

\subsection{Yang-Mills theories in $2+1$ dimensions} 
\label{2+1_Yang_Mills_subsection}

$\SU(N)$ gauge theories in $D=2+1$ spacetime dimensions are particularly interesting models characterized by non-trivial dynamics and sharing many qualitative features with ordinary Yang-Mills theories in $D=3+1$. They can be defined in terms of the following Euclidean functional integral:
\eq
\label{continuum_formulation}
Z = \int\mathcal{D} A e^{-\Se}, \;\;\; \Se = \int {\dd}^3x \frac{1}{2 g_0^2 }\Tr F_{\alpha\beta}^2,
\en
where the bare square gauge coupling $g_0^2$ has the dimensions of a mass. Thus the natural expansion parameter for bare perturbation theory calculations at a momentum scale $k$ is the dimensionless ratio $g_0^2/k$, which leads to a non-trivial infrared structure~\cite{3d_YM_renormalization_properties}. Similarly to Yang-Mills theories in $D=3+1$ dimensions, these theories are asymptotically free at high energy; one difference w.r.t. the four-dimensional case, however, is that Yang-Mills theories in $D=2+1$ dimensions are superrenormalizable (rather than renormalizable), i.e. they have a finite number of ultraviolet-divergent Feynman diagrams. 

At low energies, these theories are linearly confining with a finite mass gap and a discrete spectrum. At zero (and low) temperature, 
the physical states are color-singlet glueball states which can be classified according to the irreducible representations of the $O(2)$ 
group and charge conjugation properties (for $N>2$).

Like in the $D=3+1$ case, $\SU(N)$ Yang-Mills models in $D=2+1$ dimensions undergo a deconfinement transition at a finite critical temperature $T_c$, at which the $\Z_N$ center symmetry of the low-temperature vacuum gets spontaneously broken by a non-vanishing expectation value of the Polyakov loop in the thermodynamic limit. Again, similarly to what happens in $D=3+1$~\cite{Panero:2009tv}, the deconfinement transition is of second order for ``small'' gauge groups, and turns into a more and more strongly first-order one when the gauge group size is increased---in agreement with the intuitive picture of a more and more ``violent'' transition when a larger number of gluons get liberated~\cite{stronger_first_order_conjecture}. In particular, the transition in $D=2+1$ dimensions is of second order for the gauge groups $\SU(2)$ and $\SU(3)$, while it is a weakly first-order one for $\SU(4)$, and a stronger first-order one for $\SU(N \ge 5)$~\cite{SUN_thermodynamics_in_2_plus_1_dimensions, Liddle:2008kk} (whereas in $D=3+1$ the transition is of second order only for the $\SU(2)$ gauge group).
Here we focus our attention on the cases of $N=2$, $3$ and $4$ colors, and study these theories by regularizing them on a lattice, as discussed in the following subsection.

\subsection{Lattice regularization}
\label{SUN_lattice_regularization_subsection}

We focus our attention on the cases of $N=2$, $3$ and $4$ colors and study these theories non-perturbatively by regularizing them on a finite, isotropic cubic lattice $\Lambda$ of spacing $a$ and volume $V=L_s^2 \times L_t= (N_s^2  \times N_t )a^3$. The lattice dynamics is defined by the standard Wilson lattice gauge action:
\eq
\label{Wilson_lattice_gauge_action}
\Sel= \beta \sum_{x \in \Lambda} \sum_{1 \le \alpha < \beta \le 3} \left[1 - \frac{1}{N} \re\Tr U_{\alpha\beta}(x)\right], \;\;\; \mbox{with:} \;\;\; \beta=\frac{2N}{g_0^2 a},
\en
and:
\eq
\label{plaquette}
U_{\alpha\beta}(x) = U_\alpha(x) U_\beta(x+a\hat\alpha) U^\dagger_\alpha(x+a\hat\beta) U^\dagger_\beta(x).
\en
Periodic boundary conditions are imposed all directions, so that the shortest lattice size $L_t$ is related to the temperature of the system via: $T=1/L_t$. The functional integral appearing in eq.~(\ref{continuum_formulation}) is then replaced by the finite-dimensional multiple integral:
\eq
\label{lattice_partition_function}
\Zlat = \int \prod_{x \in \Lambda} \prod_{\alpha=1}^3 {\dd}U_\alpha(x) e^{-\Sel}
\en
(where ${\dd}U_\alpha(x)$ denotes the Haar measure for the $U_\alpha(x)$ link matrix), while expectation values of gauge-invariant physical observables $O$ are obtained from:
\eq
\label{expectation_value}
\langle O \rangle = \frac{1}{\Zlat} \int \prod_{x \in \Lambda} \prod_{\alpha=1}^3 {\dd}U_\alpha(x) \; O \; e^{-\Sel}
\en
and are estimated numerically by Monte Carlo sampling over a set of $\{ U_\alpha(x) \}$ configurations. In the following, we denote the number of configurations as $\nconf$. In our simulations, the samples of thermalized, independent configurations were produced using a $1+3$ combination of heat-bath~\cite{heatbath} and overrelaxation updates~\cite{overrelaxation} on $\SU(2)$ subgroups~\cite{Cabibbo:1982zn}.

Although high-order lattice perturbation theory computations are also available, 
typically numerical lattice simulations (both in $D=3+1$ and in $D=2+1$ dimensions) are run in a regime of strongly coupled dynamics, where the determination of the spacing $a$ as a function of the bare gauge coupling (or, equivalently, of $\beta$) is done non-perturbatively, extracting $a$ through the lattice determination of some reference quantity relevant for low-energy scales (such as, for example, the zero-temperature string tension $\sigma_0$, or the critical deconfinement temperature). The systematic uncertainty related to the choice of a physical observable to set the scale has quantitatively modest impact $\mathcal{O}(a^2)$. In our computations, we used the scale determination from ref.~\cite{Caselle:2004er} for $\SU(2)$, from ref.~\cite{Bialas:2008rk} for $\SU(3)$, and from ref.~\cite{Liddle:2008kk} for $\SU(4)$, yielding:
\eq
\label{sigma_a2}
\sigma_0 a^2 \simeq \left\{ 
\begin{array}{ll} \left[ \frac{1.324(12)}{\beta} + \frac{1.20(11)}{\beta^2} \right]^2 & \mbox{for $\SU(2)$}, \\
\left[ \frac{3.37(1)}{\beta} + \frac{3.90(25)}{\beta^2} + \frac{50.1(1.8)}{\beta^3}\right]^2 & \mbox{for $\SU(3)$}, \\
\left[ \frac{6.491(68)}{\beta} + \frac{3.7(3.2)}{\beta^2} + \frac{533(37)}{\beta^3}\right]^2 & \mbox{for $\SU(4)$}. 
\end{array}
\right.
\en 
Just to get an intuitive picture in terms of scales relevant for real-world QCD, eq.~(\ref{sigma_a2}) can be translated into a scale for $a$ in fm, by defining $\sigma_0$ to be equal to $(440~\mbox{MeV})^2$.

Table~\ref{SUN_simulation_information_table} summarizes the basic technical information about our simulations. 

\begin{table}[h]
\centering
\phantom{-------}
\begin{tabular}{|c|c|c|c|c|c|c|c|}  
\hline
 gauge group & $N_s^2 \times N_t$ & $\beta$ & $\sigma_0 a^2$ & $a$ & physical volume &  $T/T_c$ & $\nconf$ \\
\hline
$\SU(2)$ & $120^2 \times 8$ & $9.0$  & $0.0262(1)$ & $0.072$~fm & $8.7^2 \times 0.58 $~fm$^3$ &  $3/4$ & $5.6 \times 10^4$ \\
$\SU(3)$ & $120^2 \times 8$ & $21.34$  & $0.0293(1)$ & $0.077$~fm & $9.2^2 \times 0.61 $~fm$^3$ &  $3/4$ & $8 \times 10^4$ \\
$\SU(4)$ & $120^2 \times 8$ & $40.10$  & $0.0282(6)$ & $0.075$~fm & $9.0^2 \times 0.60 $~fm$^3$ &  $3/4$ & $4 \times 10^4$ \\
 \hline
\end{tabular}
\phantom{-------}
\caption{Parameters of our $\SU(N)$ lattice simulations (see the text for the definition of the various quantities).}
\label{SUN_simulation_information_table}
\end{table}

\subsection{$\Z_2$ lattice gauge theory in three dimensions}
\label{Z2_subsection}

As mentioned in the Introduction, we also compared the results from Yang-Mills theory with those from the Abelian $\Z_2$ gauge theory in three dimensions: this is a lattice model whose fundamental degrees of freedom are $\sigma_\alpha(x)$ variables defined on the oriented bonds of a three-dimensional cubic grid with periodic boundary conditions and taking values in (the fundamental representation of) the cyclic group of order two: $\Z_2 \sim \left( \{ 1 , -1 \} , \cdot \right)$. In analogy to eq.~(\ref{Wilson_lattice_gauge_action}), the dynamics of this model is defined by the action:
\eq
\label{Z_2_Wilson_lattice_gauge_action}
S_{\tiny\mbox{gauge}}= - \betagauge \sum_{x \in \Lambda} \sum_{1 \le \alpha < \beta \le 3}  \sigma_{\alpha\beta}(x), \;\;\; \mbox{with:} \;\;\; 
\sigma_{\alpha\beta}(x) = \sigma_\alpha(x) \sigma_\beta(x+a\hat\alpha) \sigma_\alpha(x+a\hat\beta) \sigma_\beta(x),
\en
which is invariant under $\Z_2$ gauge transformations flipping the signs of the six $\sigma_\alpha(x)$ variables on the bonds touching the site $x$.

Despite its deceivingly simple definition, this model is characterized by highly non-trivial dynamics and a rich phase structure. For values of $\betagauge$ less than $\beta_{\tiny\mbox{gauge, crit.}}=0.76141346(6)$~\cite{Deng_Bloete} (and for sufficiently large lattice sizes), the model is in a confining phase, while above that critical value, it is in a deconfining phase. The bulk phase transition separating the two phases is a second order one, which enables one to investigate an effective continuum limit of this confining lattice model by approaching $\beta_{\tiny\mbox{gauge, crit.}}$ from below (by virtue of the fact that $\sigma_0 a^2 \to 0$ for $\betagauge \to {\beta^{-}_{\tiny\mbox{gauge, crit.}}}$) and scaling the lattice size appropriately. Furthermore, in the confining phase there also exists an infinite-order roughening transition at $\beta_{\tiny\mbox{gauge, rough.}}=0.47542(1)$~\cite{Hasenbusch_Pinn}, separating the  $\betagauge<\beta_{\tiny\mbox{gauge, rough.}}$ regime where strong-coupling expansions hold, from a ``rough'' phase $\beta_{\tiny\mbox{gauge, rough.}}<\betagauge<\beta_{\tiny\mbox{gauge, crit.}}$ in which the theory can support massless excitations. Furthermore, when this model is defined on a lattice with a compact dimension of size $L_t$, one can observe a finite-temperature deconfinement transition when $L_t$ becomes shorter than the inverse of the critical temperature. 

An interesting feature of the $\Z_2$ gauge model defined by eq.~(\ref{Z_2_Wilson_lattice_gauge_action}) is that, through a duality transformation~\cite{Kramers:1941kn}, it can be mapped exactly to the tridimensional Ising spin model with $s(x) \in \Z_2$ variables, which is defined by:
\eq
\label{Ising_spin_model}
S_{\tiny\mbox{spin}}= - \betaspin \sum_{(x, y)} s(x) s(y), \;\;\; \mbox{with:} \;\;\; \sinh(2\betaspin)\sinh(2\betagauge)=1,
\en
where $(x, y)$ denotes that the sites $x$ and $y$ are nearest-neighbors. 
This duality implies a one-to-one mapping between the free energy densities of the two models in the thermodynamic limit, and allows one to express expectation values of
gauge-invariant observables in the $\Z_2$ gauge model as ratios of partition functions of the spin system, with a set of appropriately chosen \emph{antiferromagnetic}
couplings, leading to a major efficiency boost in simulations of this model.

In particular, in this work we studied the $\Z_2$ gauge model analyzing two ensembles of configurations corresponding to two different temperatures $T=T_c/2$ and $T=2T_c/3$, taken from ref.~\cite{Caselle:2002ah}; technical details are summarized in table~\ref{Z2_simulation_information_table}.

\begin{table}[h]
\centering
\phantom{-------}
\begin{tabular}{|cc|cc|}  \hline
\multicolumn{2}{|c|}{Ensemble I} & \multicolumn{2}{|c|}{Ensemble II} \\
\hline
$N_s^2 \times N_t$ & $96^2 \times 12$  & $N_s^2 \times N_t$ & $96^2 \times 9$ \\
$\betagauge $ & $0.746035$ & $\betagauge $ & $0.746035$ \\
$\sigma_0 a^2$ & $0.018943$ & $\sigma_0 a^2$ & $0.018943$ \\
$a$ & $0.062$~fm & $a$ & $0.062$~fm \\
physical volume & $5.92^2 \times 0.74 $~fm$^3$ & physical volume & $5.92^2 \times 0.55$~fm$^3$ \\
$T/T_c$ & $1/2$ & $T/T_c$ & $2/3$ \\
$\nconf$ & $3.2 \times 10^9$ & $\nconf$ & $3.2 \times 10^9$ \\
\hline
\end{tabular}
\phantom{-------}
\caption{Parameters of the $\Z_2$ simulations analyzed in this work.}
\label{Z2_simulation_information_table}
\end{table}

\subsection{Observables}
\label{observables_subsection}

In order to investigate the logarithmic term appearing in eq.~(\ref{potential_approximate_form}), we estimate numerically the expectation values of (on-axis) two-point correlation functions of Polyakov loops at a finite temperature $T=1/(a N_t) < T_c$:
\eq
\label{lattice_Polyakov_correlator}
\Glat(r,T) = \frac{1}{\Zlat} \int \prod_{x \in \Lambda} \prod_{\alpha=1}^3 {\dd}U_\alpha(x) \mathcal{P}^*(a r, 0) \mathcal{P}(0, 0) \; e^{-\Sel} \;,
\en
with:
\eq
\label{Polyakov_loop_definition}
\mathcal{P}(an_x, an_y) = \Tr \prod_{n_t=0}^{N_t-1} U_t ( an_x, an_y, an_t).
\en
To measure $\Glat(r,t)$, we used a highly efficient multi-level algorithm~\cite{Luscher:2001up}, which allows one to achieve exponential error-reduction. 
The error analysis was done using the standard jackknife method~\cite{jackknife}, taking cross-correlations into account. 

Following ~\cite{Luscher:2002qv} and~\cite{Caselle:2004er}
it is particularly convenient to introduce the following quantities:
\eqar
\label{Q_definition}
Q(r,T) &=& \frac{T}{a} \ln \frac{\Glat(r,T)}{\Glat(r+a,T)}, \\
\label{A_definition}
A(r,T) &=& \frac{r^2}{a^2} \ln \frac{\Glat(r+a,T)\Glat(r-a,T)}{\Glat^2(r,T)}.
\enar
Note that, in the continuum limit $a \to 0$, $Q(r,T)$ tends to the first derivative of $V(r,T)$ w.r.t. $r$:
\eq
\label{continuum_limit_of_Q}
\lim_{a \to 0} Q = \frac{\partial V}{\partial r},
\en
so that it can be interpreted as a lattice version of (minus) the interquark force. On the other hand, $A(r,T)$ is a dimensionless quantity proportional to the discretized derivative of the force:
\eq
\label{continuum_limit_of_A}
\lim_{a \to 0} A = - \frac{r^2}{T} \frac{\partial^2 V}{\partial r^2}.
\en

It is easy to see that these quantities are the finite temperature version of the observables introduced in~\cite{Luscher:2002qv}. In particular
 $Q(r,T)$ coincides in the low-$T$ limit with the ``force'' $F(R)$ of~\cite{Luscher:2002qv} 
while $A(r,T)$ is related to the ``central charge'' $c(R)$  of~\cite{Luscher:2002qv}  as follows
\eq
A(r,T)=\frac{2}{rT}c(r).
\en

Using eq.~(\ref{G_dominated_by_K0}), eq.~(\ref{T-dependent_sigma}) and eq.~(\ref{K0_expansion}) we may estimate the large-$r$ limit of 
these two observables in the framework of the
Nambu-Goto effective string model:
\eqar
\label{NG_prediction_for_Q} Q(r,T) &\simeq& \sigma + \frac{T}{2r} - \frac{T^2}{8 \sigma r^2}, \\
\label{NG_prediction_for_A} A(r,T) &\simeq& \frac{1}{2} - \frac{T}{4 \sigma r}.
\enar

In these two functions the universal logarithmic term in which we are interested is encoded in the coefficient $1/(2r)$ 
of the term proportional to $T$ in eq.~(\ref{NG_prediction_for_Q}) and in the constant $1/2$ in eq.~(\ref{NG_prediction_for_A}). This last value, 
thanks to the normalization of $A(r,T)$ 
in eq.~(\ref{A_definition}), exactly coincides with the coefficient of the logarithmic term in the interquark potential. 
In the next section we shall use our Monte Carlo simulations to estimate these quantities in the $\SU(N)$ and $\Z_2$ models that we studied.

\section{Results}
\label{results_section}

   \subsection{General remarks on the fitting procedure}
Let us briefly describe a few features of our fitting procedure.
\begin{itemize}
\item
As previously discussed, the effective string description holds for large enough interquark distances. 
Following previous works on the finite temperature interquark potential, we decided to include in the fit only values of the interquark distance $r>\rmin$ such that 
$\sigma(T) \rmin^2 \sim 1.5$. To fulfill this constraint we had to choose $\rmin=9a$ for $\SU(N)$ models,  $\rmin=10a$ for Ising, Ensemble I and  $\rmin=12a$ 
for Ising, Ensemble II. As a consistency check, it is interesting to notice that with this choice in all the cases the reduced $\chi^2$ turned out to be always 
of order unity.
\item
  In the $\SU(N)$ case, the data for $Q$ turned out to be strongly cross-correlated and thus we had to perform our fits taking the cross-correlation matrix into account.   
 It is important to stress that 
   a naive fit to the
   data neglecting cross-correlations would lead not only to different values of the reduced $\chi^2$ and of the statistical uncertainties
   but also to differences in the best fit values larger than the quoted errors. 
   On the contrary, the Ising data for different values of $r$ are completely uncorrelated since they were obtained in different simulations.
\end{itemize}

   \subsection{Analysis of the $ \SU(2) $ data}

Table~\ref{SU2_results_table} reports our numerical results for the ``discretized'' interquark force $Q(r,T)$ defined by eq.~(\ref{Q_definition}), 
as obtained from $\SU(2)$ simulations at fixed $T= 3T_c/4 $. 
For convenience, we express both $r$ and $Q$ in units of (the appropriate power of) the lattice spacing $a$.

\begin{table}[ht]
\centering
\phantom{-------}
\begin{tabular}{|cc|cc|cc|}
\hline
 $r/a$ & $a^2 Q$ &  $r/a$ & $a^2 Q$ &  $r/a$ & $a^2 Q$ \\
\hline
   2 & 0.037433(46)  &      8 & 0.02232(11)   &        14 & 0.01971(16)   \\
   3 & 0.030958(56)  &      9 & 0.02170(12)   &        15 & 0.01949(18)    \\
   4 & 0.027600(64)  &     10 & 0.02117(12)   &        16 & 0.01926(19)    \\
   5 & 0.025553(72)  &     11 & 0.02072(13)   &        17 & 0.01906(20)    \\
   6 & 0.024154(84)  &     12 & 0.02034(15)   &        18 & 0.01892(22)    \\
   7 & 0.023118(94)  &     13 & 0.02000(15)   &        19 & 0.01876(24)    \\
\hline
\end{tabular}
\phantom{-------}
\caption{Results for $a^2 Q(r,T)$, as a function of the interquark distance $r/a$, from the  $\SU(2)$ simulations at $T=3T_c/4$.}
\label{SU2_results_table}
\end{table}

   We fitted the values of $ Q(r,T) $ according to
\eq
\label{SU2_Q_fit}
a^2 Q (r,T)|_{T=3T_c/4} = s + \frac{b}{r/a} + \frac{c}{(r/a)^2} \, , 
\en
   and found the following best fit values for the parameters
   \begin{equation} \nonumber
    s= 0.01530(37)\ 
    b= 0.0668(58)\
    c= - 0.087(27)
   \end{equation}
   with a reduced $\chi^2_r=0.75$.

   The universal correction in which we are interested is encoded in the parameter $b$ which according to the analysis discussed in the previous sections should be given
   by
   \begin{equation} \nonumber
    b=\frac{aT}{2}=\frac{1}{16}=0.0625 \, ,
   \end{equation}
   which turns out to be in remarkable agreement with our result.
   
   This is further confirmed by the analysis of the $A(r,T)$ values (which can be easily obtained from the data reported
   in table~\ref{SU2_results_table}). We fitted these values to
\eq
\label{SU2_A_fit}
A (r,T)|_{T=3T_c/4} = k - \frac{m}{r/a} \, , 
\en
   finding
   \begin{equation} \nonumber
    k=0.528(28),\;\; 
    m=-1.09(28), \;\;\; \mbox{with  $\redchisq=1.6$},
   \end{equation}
   which is again in perfect agreement with the expected value $k=1/2$. 
   
   From the first fit we can extract the value $\sigma(T)=0.01530(37)$ for the finite temperature string tension at $aT=1/8$.  
   Using the value $\sigma_0 a^2$ reported in eq.~(\ref{sigma_a2}), we may obtain a ``Nambu-Goto'' prediction for $\sigma(T)$ using 
   eq.~(\ref{T-dependent_sigma}), which turns out to be
   $a^2\sigma_{\mbox{\tiny{NG}}}\left(aT=1/8\right)=0.01605(6)$, at two standard deviations from the observed value. 
   This indicates, as already observed in \cite{Athenodorou:2007du},
   that for the (2+1) $\SU(2)$ LGT the Nambu-Goto string represents a rather good approximation but, with the new generation of high precision data, small deviations start to
   be detectable.
   
   Finally, in the same way, we may obtain predictions for the subleading corrections in the two fits.
   We find for the $c$ term in the first fit    $c_{\mbox{\tiny{NG}}} \sim -0.1216(5)$ and for $m$ in the second fit $m_{\mbox{\tiny{NG}}}= -1.946(8)$. Both values are of the same 
   order of magnitude and of the same sign of the  results, but they are not
   compatible within the errors. 
   This discrepancy agrees in sign and magnitude with the analogous deviations from the Nambu-Goto ansatz observed in \cite{Athenodorou:2007du} and summarized in the 
   coefficient $C_3$ evaluated in \cite{Athenodorou:2007du}.

   A similar pattern is observed in the $\SU(3)$ and $\SU(4)$ cases, which we describe in detail below.

   \subsection{Analysis of the $\SU(3)$ data}

For the $\SU(3)$ gauge group, our results for $a^2 Q(r,T)$, as a function of $r/a$, are reported in table~\ref{SU3_results_table}.

\begin{table}[ht]
\centering
\phantom{-------}
\begin{tabular}{|cc|cc|cc|}  
\hline
 $r/a$ & $a^2 Q$ &  $r/a$ & $a^2 Q$ &  $r/a$ & $a^2 Q$ \\
\hline
   2 &  0.042353(20)   &      8 &  0.025525(47)  &        14 &   0.022881(96) \\
   3 &  0.035013(24)   &      9 &  0.024857(54)  &        15 &   0.02262(11)   \\
   4 &  0.031241(28)   &     10 &  0.024326(61)  &        16 &   0.02238(12)   \\
   5 &  0.028971(33)   &     11 &  0.023876(68)  &        17 &   0.02223(14)   \\
   6 &  0.027447(37)   &     12 &  0.023501(76)  &        18 &   0.02196(15)   \\
   7 &  0.026351(42)   &     13 &  0.023184(88)  &        19 &   0.02178(17)   \\
\hline
\end{tabular}
\phantom{-------}
\caption{Results for $a^2 Q(r,T)$, as a function of the interquark distance $r/a$, from the  $\SU(3)$ simulations; the temperature is fixed at $T=3T_c/4$.}
\label{SU3_results_table}
\end{table}

In this case, the fit to eq.~(\ref{NG_prediction_for_Q}) yields
\eq
\label{SU3_Q_fit}
a^2 Q (r,T)|_{T=3T_c/4} = 0.01884(44) + \frac{0.0612(74)}{r/a} - \frac{0.063(34)}{(r/a)^2}, \;\;\; \mbox{with  $\redchisq=1.6$},
\en
to be compared with $0.01946(6)$, $0.0625$ and $-0.1003(5)$ from the Nambu-Goto model, 
while the fit of $A(r, T)$ gives
\eq
\label{SU3_A_fit}
A (r,T)|_{T=3T_c/4} = 0.529(46) - \frac{1.00(45)}{r/a}, \;\;\; \mbox{with  $\redchisq=0.9$},
\en
to be compared with $k=1/2$  and $m_{\mbox{\tiny{NG}}}=-1.605(6)$.

   \subsection{Analysis of the $\SU(4)$ data}

We report our results for $Q(r,T)$ for $\SU(4)$ gauge group in table~\ref{SU4_results_table}: they can be fitted to eq.~(\ref{NG_prediction_for_Q}) as:
\eq
\label{SU4_Q_fit}
a^2 Q (r,T)|_{T=3T_c/4} = 0.01721(43) + \frac{0.0634(70)}{r/a} - \frac{0.063(32)}{(r/a)^2}, \;\;\; \mbox{with  $\redchisq=1.4$},
\en
to be compared with $0.0183(4)$, $0.0625$ and $-0.107(2)$ from the Nambu-Goto model, 
while the fit of $A(r, T)$ gives: 
\eq
\label{SU4_A_fit}
A (r,T)|_{T=3T_c/4} = 0.470(30) - \frac{0.38(30)}{r/a}, \;\;\; \mbox{with  $\redchisq=0.7$},
\en
to be compared with $k=1/2$  and $m_{\mbox{\tiny{NG}}}=-1.71(4)$.

It is interesting to observe that, although the disagreement with the Nambu-Goto values increases as $N$ increases, the expectation value of  $k$ (or equivalently of
$b$ in the first fit) is in perfect agreement with the expected value $k=1/2$  (or $b=1/16$), thus showing that this 
term is indeed universal (in the same sense in which the
L\"uscher term is universal) and is a robust indicator of the free bosonic nature of the transverse excitations of the effective string.

\begin{table}[ht]
\centering
\phantom{-------}
\begin{tabular}{|cc|cc|cc|}  
\hline
 $r/a$ & $a^2 Q$ &  $r/a$ & $a^2 Q$ &  $r/a$ & $a^2 Q$ \\
\hline
   2 & 0.040834(29)  &      8 &  0.024044(58)   &        14 &  0.02142(11)  \\
   3 & 0.033492(33)  &      9 &  0.023379(65)   &        15 &  0.02118(13)   \\
   4 & 0.029726(37)  &     10 &   0.022847(73)  &        16 &  0.02093(14)   \\
   5 & 0.027468(42)  &     11 &   0.022403(81)  &        17 &  0.02068(16)   \\
   6 & 0.025953(47)  &     12 &   0.022030(91)  &        18 &  0.02052(18)   \\
   7 & 0.024871(53)  &     13 &   0.02170(10)   &        19 &  0.02037(20)   \\
\hline
\end{tabular}
\phantom{-------}
\caption{Results for $a^2 Q(r,T)$, as a function of $r/a$, from our $\SU(4)$ simulations at $T=3T_c/4$.}
\label{SU4_results_table}
\end{table}

\begin{figure*}
\centerline{\includegraphics[width=.45\textwidth]{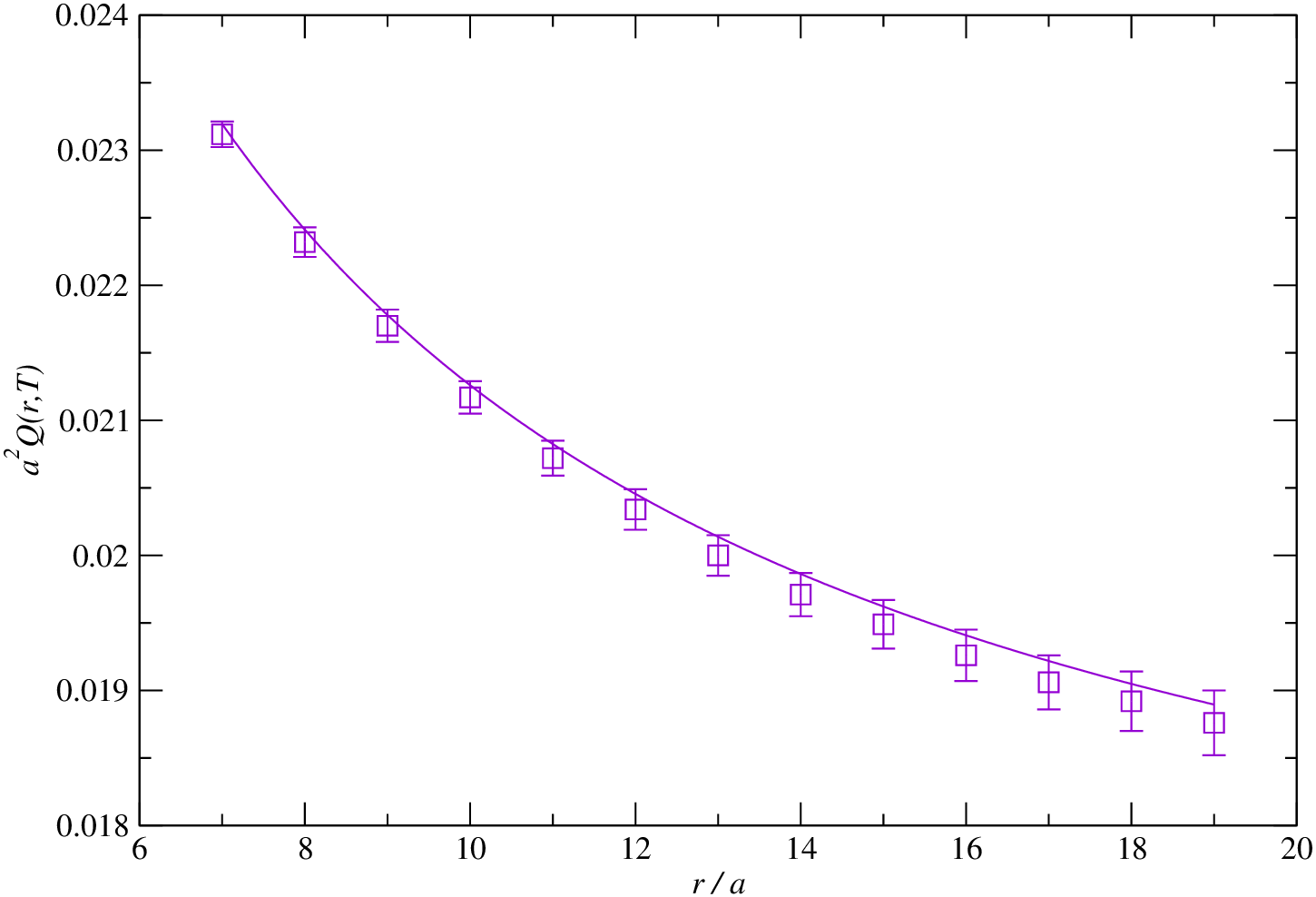} \hfill \includegraphics[width=.44\textwidth]{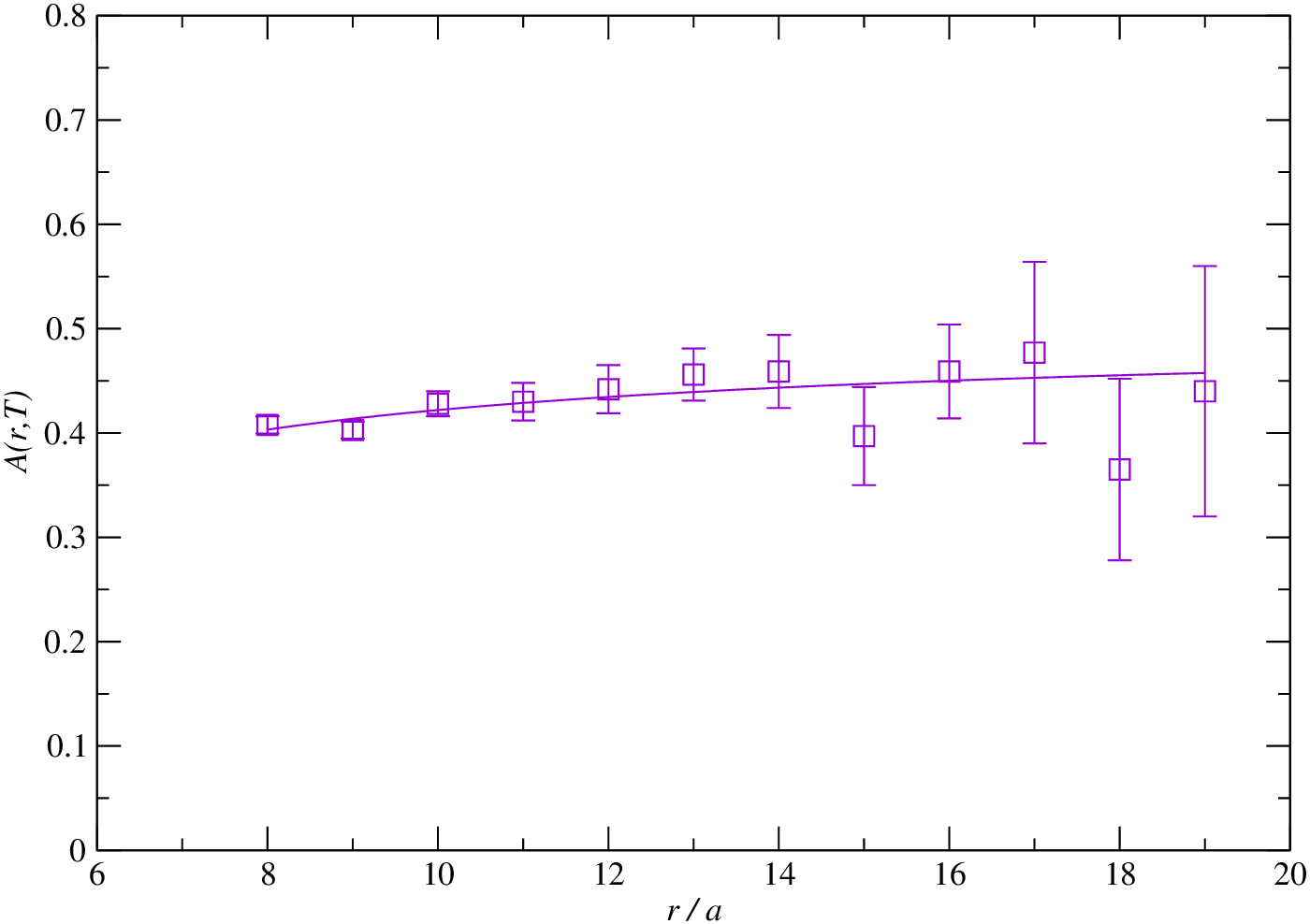}}
\vspace{1cm}
\centerline{\includegraphics[width=.45\textwidth]{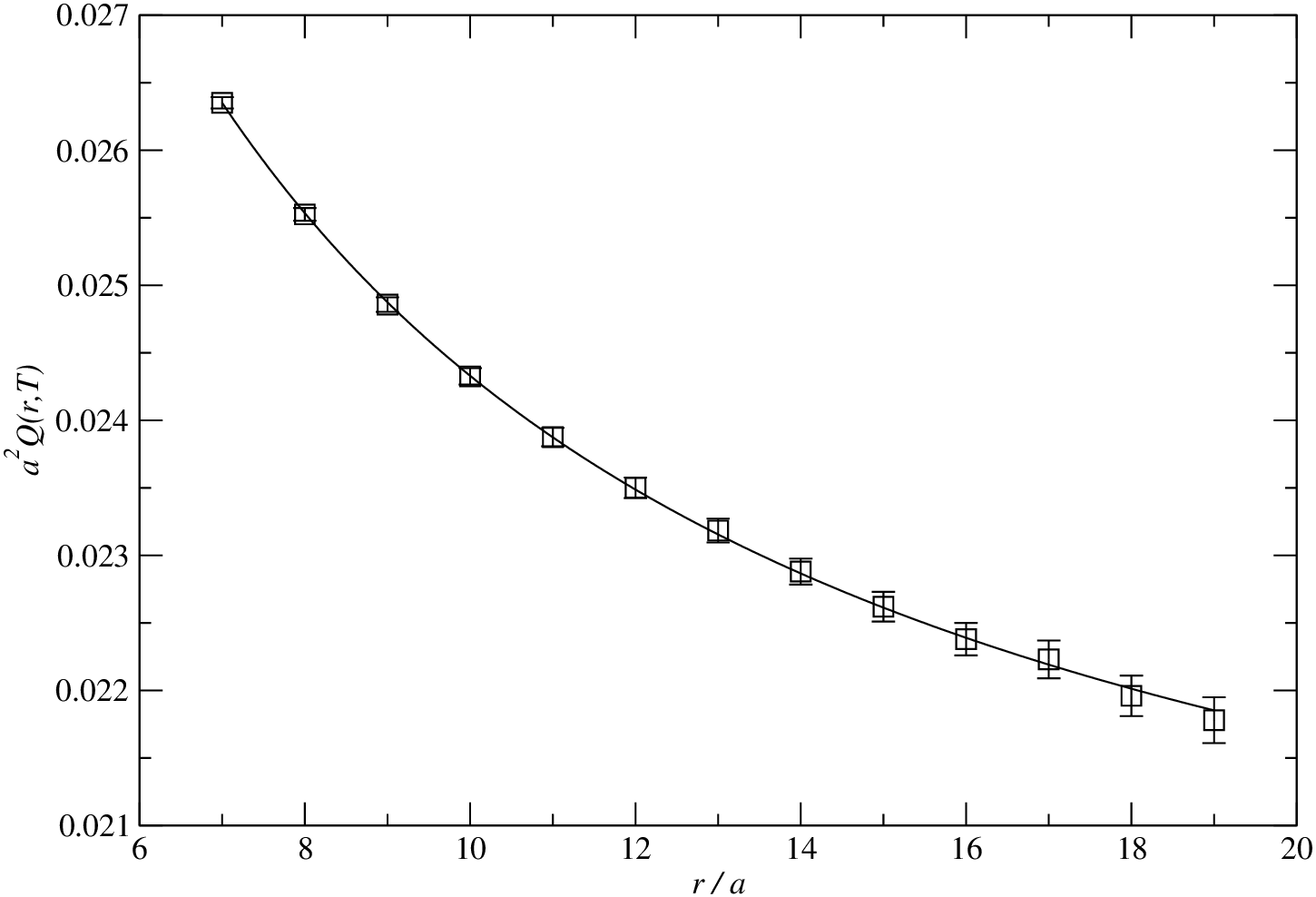} \hfill \includegraphics[width=.44\textwidth]{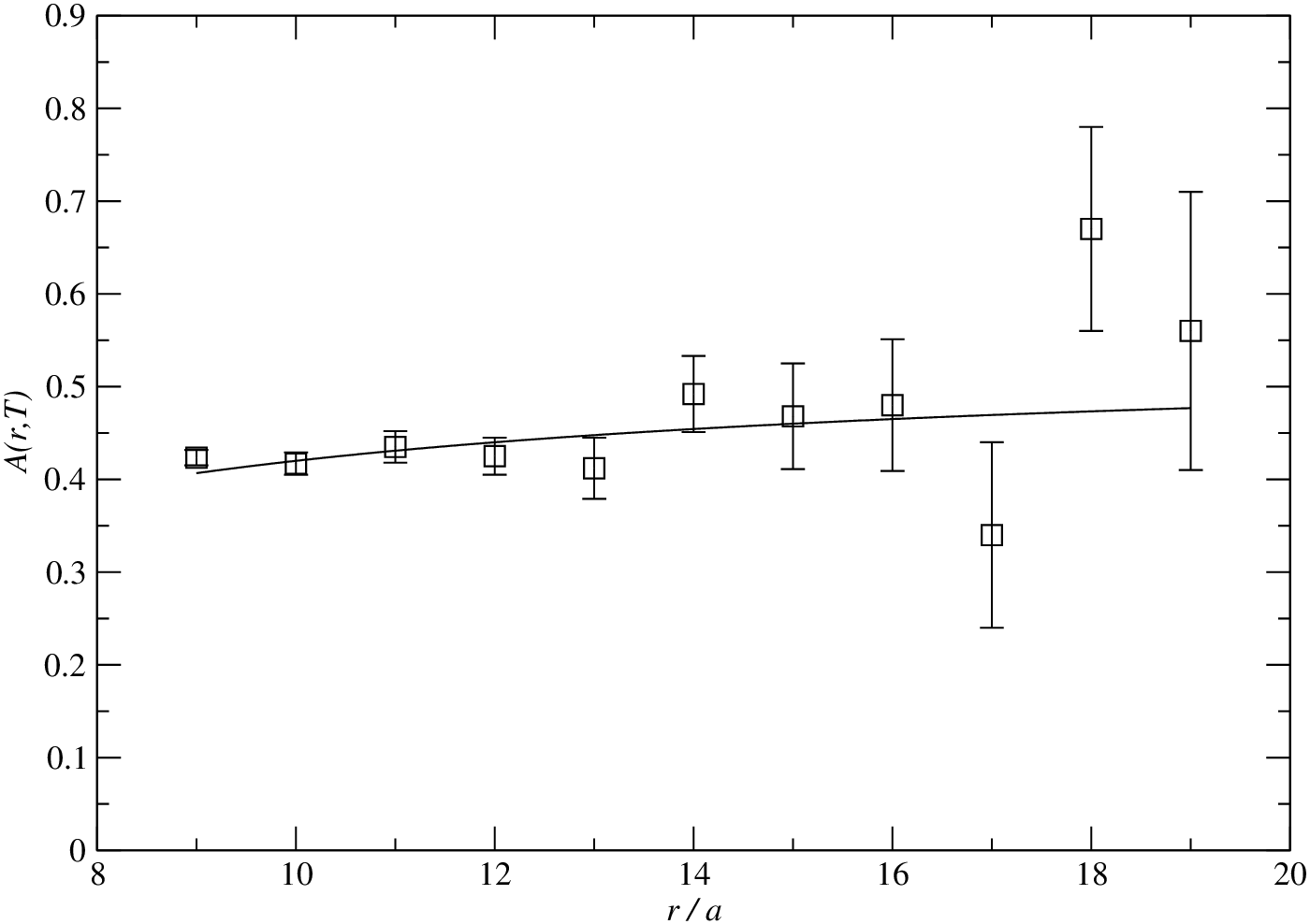}}
\vspace{1cm}
\centerline{\includegraphics[width=.45\textwidth]{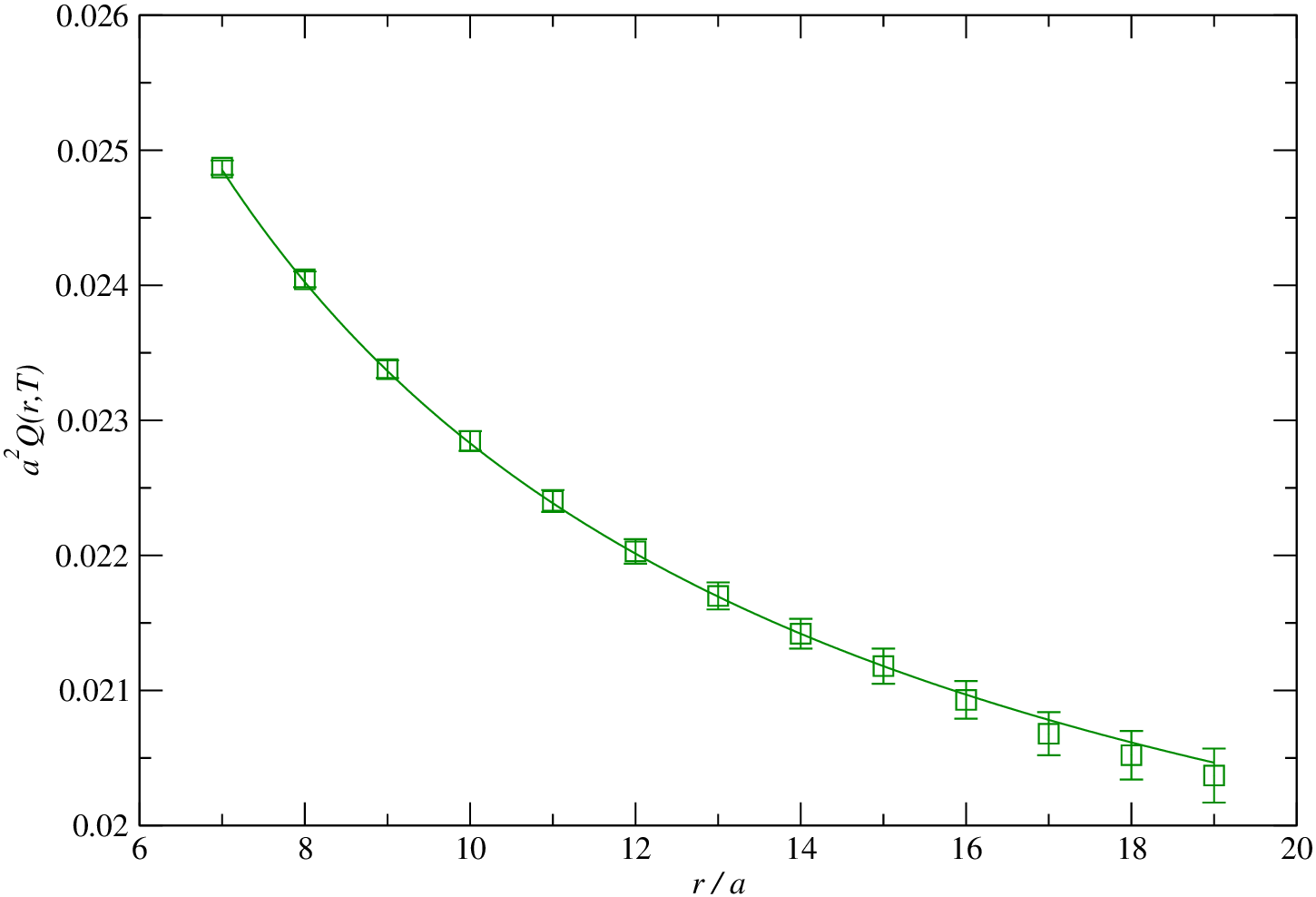} \hfill \includegraphics[width=.44\textwidth]{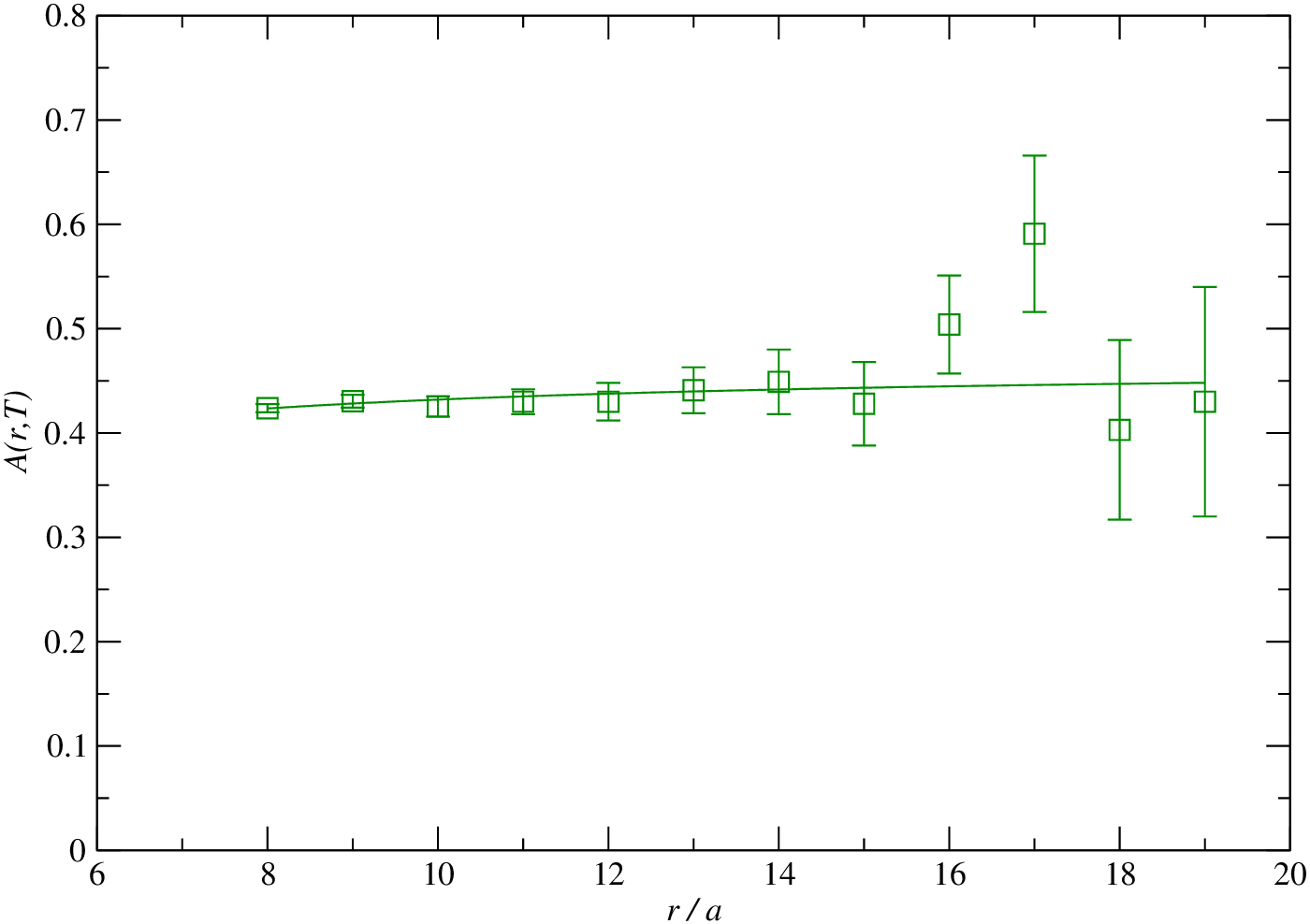}}
\caption{Left-hand-side panels: Results for the ``discretized interquark force'' $a^2 Q(r,T)$ defined in eq.~(\ref{Q_definition}), as a function of the source separation in units of the lattice spacing $r/a$, at fixed $T=3T_c/4$, together with the fitted curves, 
for $\SU(2)$ (top), $\SU(3)$ (middle) and $\SU(4)$ (bottom). Right-hand-side panels: Same, but for $A(r,T)$ as a function of $r/a$, together with the fitted curves.}
\label{SUN_results_fig}
\end{figure*}

The results for all the three gauge groups, and the fitted curves, are shown in fig.~\ref{SUN_results_fig}.

   \subsection{Analysis of the $\Z_2$ data}

As usual, the data in the Ising case show a slightly different behavior.

In the ensemble~I (which has $T/T_c=1/2$) the agreement with Nambu-Goto is remarkably good, while for the ensemble~II (which is closer to the critical point, with
$T/T_c=2/3$) sizeable deviations from the Nambu-Goto predictions can be observed. On the contrary,
the values of $k$ and $b$ are fully compatible, like for $\SU(N)$, with the
effective string expectations for both ensembles.

 Our results for $Q(r,T)$ at $T=T_c/2$ are reported in table~\ref{Z2_cold_results_table}, and can be fitted to:
\eq
\label{Z2_cold_Q_fit}
a^2 Q (r,T)|_{T=T_c/2} = 0.01485(2) + \frac{0.0414(8)}{r/a} - \frac{0.049(6)}{(r/a)^2}, \;\;\; \mbox{with  $\redchisq=0.7$},
\en
to be compared to the prediction from the Nambu-Goto model: $0.01487$, $0.04167$ and $-0.58$ (due to the precision in the determination of $\sigma_0$, errors in the
Nambu-Goto estimates of these parameters can be neglected), 
while for $A(r, T)$ the corresponding results\footnote{Note that, 
since in the $\Z_2$ simulations we determined $Q(r,T)$ only for even values of $r/a$, in this case we modified 
the definition of $A(r, T)$ in eq.~(\ref{A_definition}) by replacing $a$ with $2a$.} read: 
\eq
\label{Z2_cold_A_fit}
A (r,T)|_{T=T_c/2} = 0.47(3) - \frac{1.4(3)}{r/a}, \;\;\; \mbox{with  $\redchisq=0.85$}, 
\en
to be compared with $k=1/2$ and  $m=-1.40$.

\begin{table}[ht]
\centering
\phantom{-------}
\begin{tabular}{|cc|cc|cc|}  
\hline
 $r/a$ & $a^2 Q$ &  $r/a$ & $a^2 Q$ &  $r/a$ & $a^2 Q$ \\
\hline
    8   &   0.0192898(84)    &      18  &   0.017010(13)     &      28  &   0.016285(15)  \\ 
   10  &   0.0185110(94)    &       20  &   0.016811(13)     &      30  &   0.016179(16)  \\ 
   12  &   0.0179654(93)    &       22  &   0.016633(13)     &      32  &   0.016125(15)  \\ 
   14  &   0.017568(11)     &       24  &   0.016500(13)     &       36 &  0.015960(17)    \\
   16  &   0.017242(11)     &       26  &   0.016361(14)     &       40 &  0.015838(18)    \\
\hline
\end{tabular}
\phantom{-------}
\caption{Results for $a^2 Q(r,T)$, as a function of the interquark distance $r/a$, from the  $\Z_2$ simulations at $T=T_c/2$.}
\label{Z2_cold_results_table}
\end{table}

Similarly, table~\ref{Z2_hot_results_table} displays the results for $Q(r,T)$ for the ensemble at $T/T_c=2/3$; in this case the results of the fits are:
\eq
\label{Z2_hot_Q_fit}
a^2 Q (r,T)|_{T=2T_c/3} = 0.01137(11) + \frac{0.0522(40)}{r/a} - \frac{0.076(34)}{(r/a)^2} \;\;\; \mbox{with  $\redchisq=1.8$},
\en
for $Q(r,T)$ (for which the coefficients predicted by the string model are $0.01067$, $0.0556$ and $-0.145$, respectively), and: 
\eq
\label{Z2_cold_A_fitbis}
A (r,T)|_{T=2T_c/3} = 0.47(6) - \frac{1.9(9)}{r/a}, \;\;\; \mbox{with  $\redchisq=1.1$},
\en
for $A(r,T)$, to be compared with $k=1/2$ and  $m=-2.60$.

\begin{table}[ht]
\centering
\phantom{-------}
\begin{tabular}{|cc|cc|cc|}  
\hline
 $r/a$ & $a^2 Q$ &  $r/a$ & $a^2 Q$ &  $r/a$ & $a^2 Q$ \\
\hline
   8 & 0.016872(14) &   18 & 0.014021(21) &  28 & 0.013131(25)  \\
 10 & 0.015904(15) &   20 & 0.013755(21) &  30 & 0.013022(25)  \\
 12 & 0.015194(17) &   22 & 0.013614(21) &  32 & 0.012888(26)  \\
 14 & 0.014728(18) &   24 & 0.013460(23) &       &                          \\
 16 & 0.014320(19) &   26 & 0.013292(25) &       &                          \\
\hline
\end{tabular}
\phantom{-------}
\caption{Results for $a^2 Q(r,T)$, as a function of the interquark distance $r/a$, from computations for the $\Z_2$ model at $T=2T_c/3$.}
\label{Z2_hot_results_table}
\end{table}

\begin{figure*}
\centerline{\includegraphics[width=.45\textwidth]{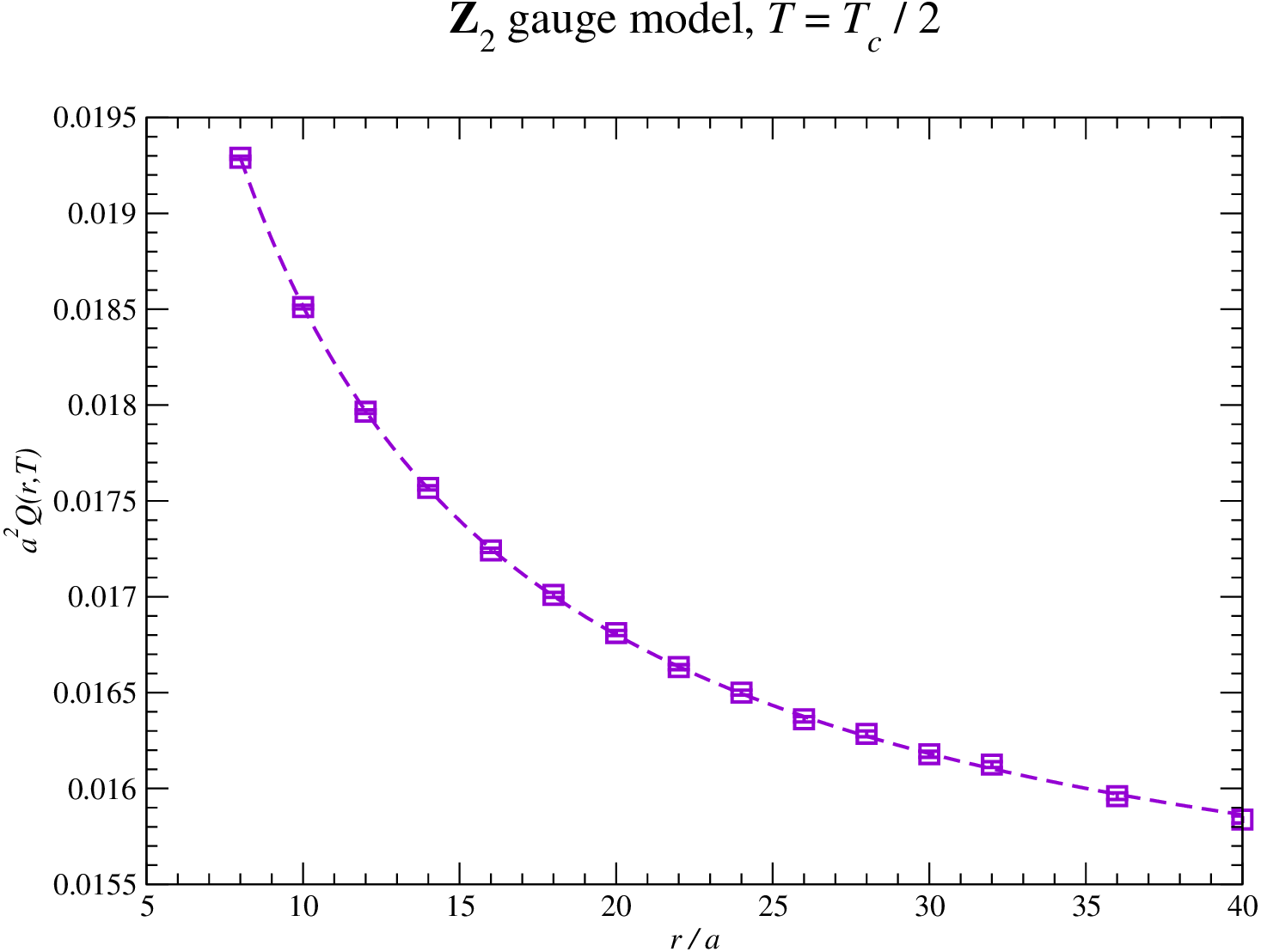} \hfill \includegraphics[width=.44\textwidth]{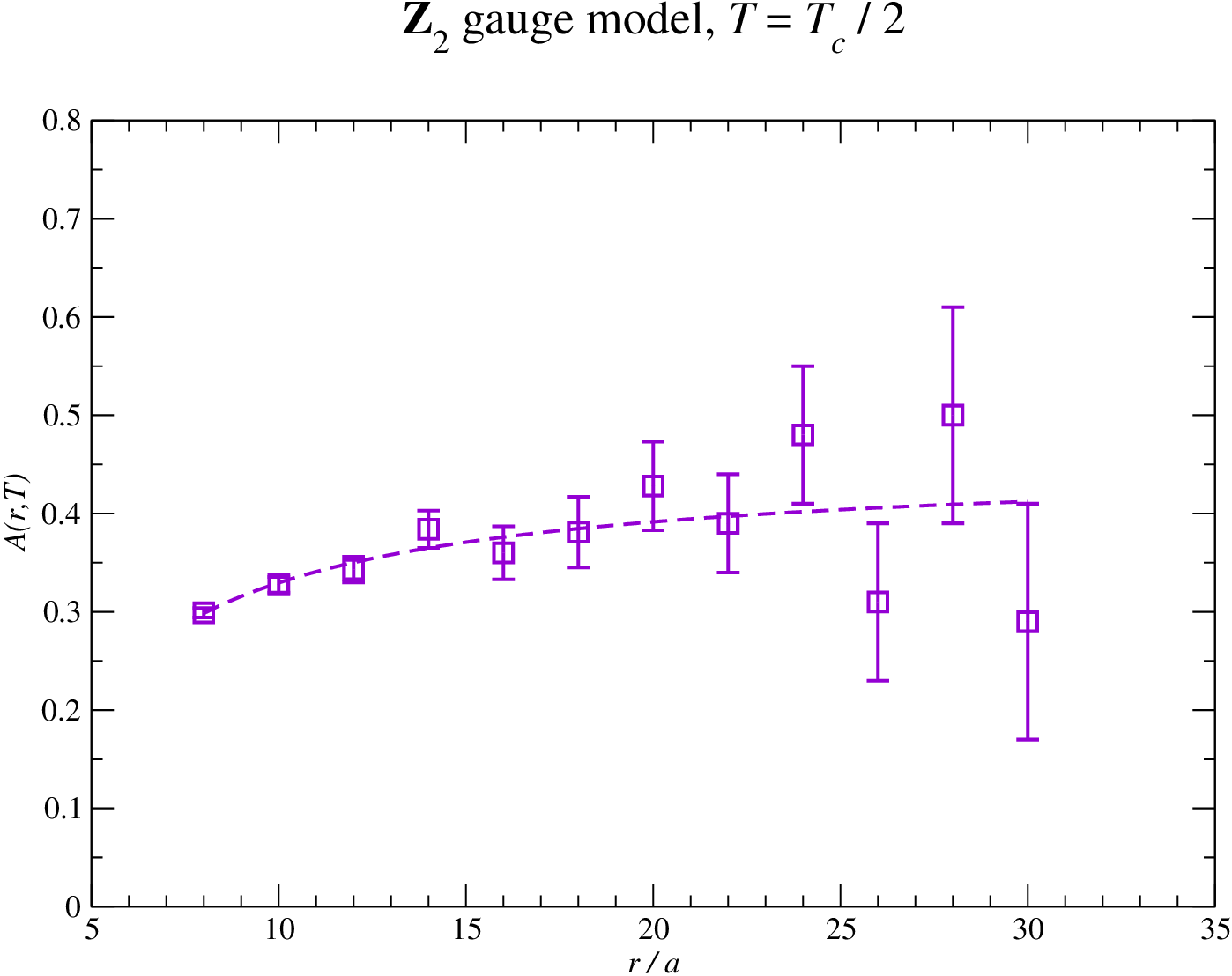}}
\vspace{1cm}
\centerline{\includegraphics[width=.45\textwidth]{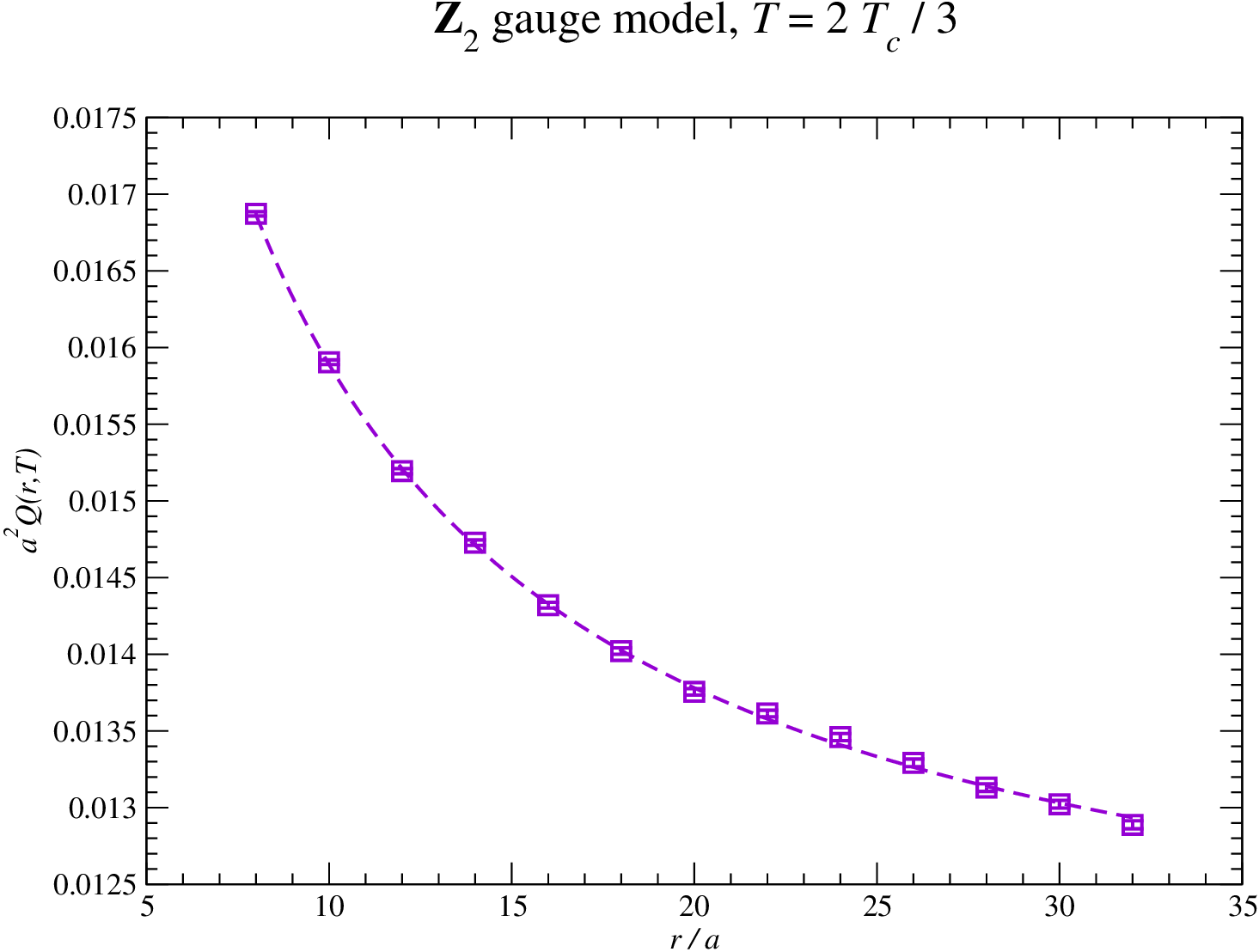} \hfill \includegraphics[width=.44\textwidth]{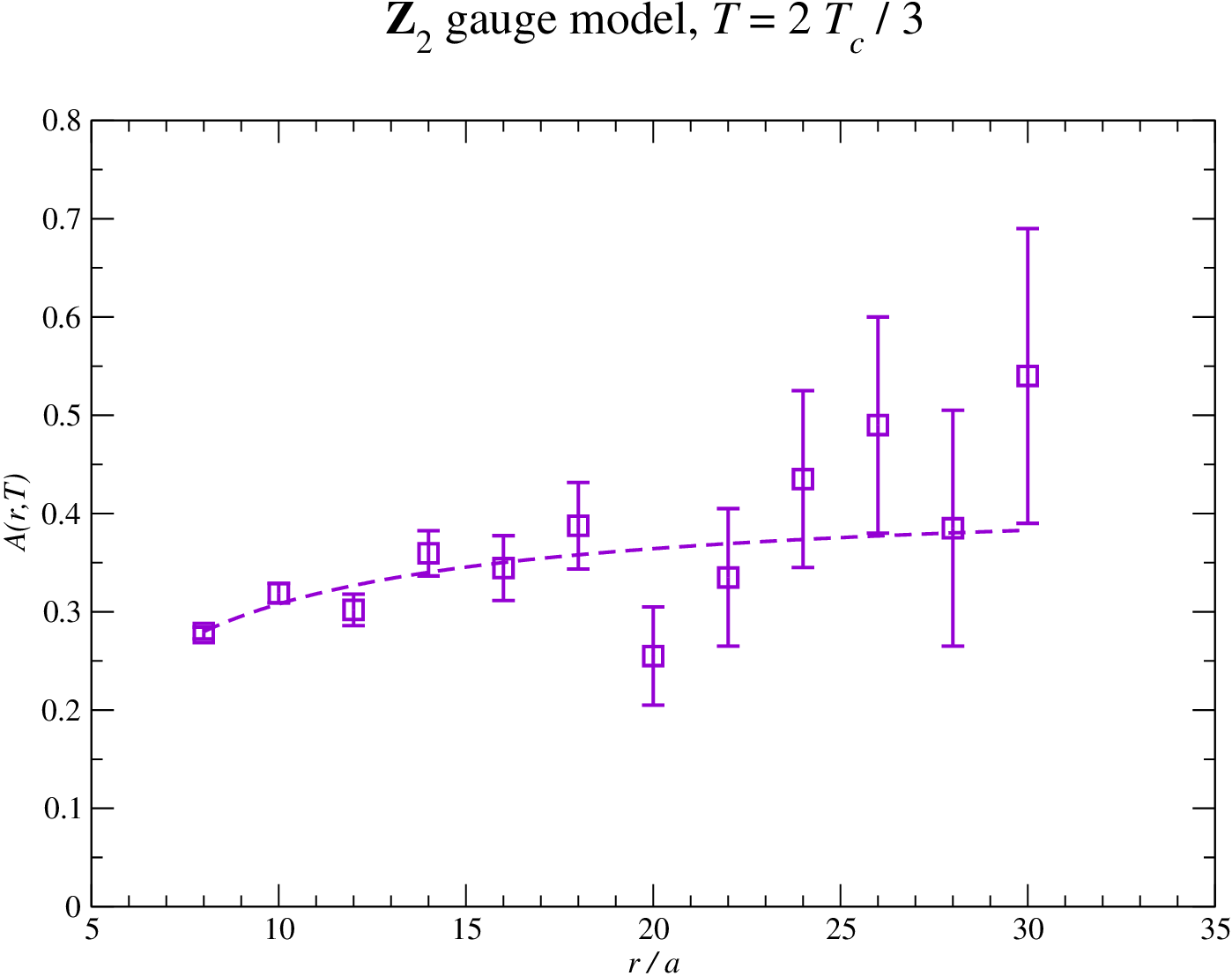}}
\caption{Same as in figure~\ref{SUN_results_fig}, but for the $\Z_2$ gauge model at $T=T_c/2$ (top) and at $T=2T_c/3$ (bottom).}
\label{Z2_results_fig}
\end{figure*}

Figure~\ref{Z2_results_fig} displays our results from the two $\Z_2$ ensembles at $T=T_c/2$ and $T=2T_c/3$, together with the fitted curves.

It is interesting to observe that (as in the $\SU(N)$ case)
the values of $k$ and $b$ are fully compatible  with the
effective string expectations for both ensembles, even though in the second ensemble the other parameters disagree within the errors from the Nambu-Goto estimates.

Finally, let us mention that, as already noticed in \cite{Caselle:2002ah,truncated}, the best fit values obtained in this second ensemble
 turn out to agree with a
truncation of the Nambu-Goto model up to the $\mathcal{O}(T^4/\sigma_0)$ term in the expansion of eq.~(\ref{T-dependent_sigma}), which gives
 $s=0.01138$.

\section{Discussion and conclusions}
\label{conclusions_section}

The results presented in the previous section show that, in all cases that we studied,
 the numerical data are in  good (but not exact) agreement with the predictions from the Nambu-Goto 
string model, eq.~(\ref{NG_prediction_for_Q}) and eq.~(\ref{NG_prediction_for_A}). This is particularly true for the $\SU(N)$ models, while
in the Ising case one observes increasing deviations as the temperature increases. 

However in all the cases, even when the Nambu-Goto prediction for the string tension is not compatible within the error with the first
term of eq.~(\ref{NG_prediction_for_Q}), the coefficients of the logarithmic term in the potential (which is encoded in the $b$ 
term of eq.~(\ref{NG_prediction_for_Q}) or, equivalently, in the $k$ term of eq.~(\ref{NG_prediction_for_A}))
always agree with the predicted value. 
In particular, the dimensionless ratio $A(r,T)$ at fixed $T$ clearly tends to 
the expected value $k=1/2$ in the limit of large quark-antiquark separations $r \to \infty$.

The fact that $A$ tends to the asymptotic value $1/2$ indicates that, whatever the exact effective string action 
describing the world-sheet dynamics 
in this temperature range, its large-$r$ limit is represented by a $c=1$ conformal field theory describing a free \emph{uncompactified} 
bosonic degree of freedom. In this respect, this logarithmic term can be considered as a high-temperature analogue of the L\"uscher term and,
exactly as the L\"uscher term, it is universal (as we have seen in our simulations) and has a simple linear dependence on the number
of transverse dimensions. 

As mentioned in the introduction, the L\"uscher term in the high temperature regime, due to the peculiar modular symmetry of the string partition function,
becomes a correction proportional to the interquark distance. The same happens to all the higher order corrections due to the higher order terms in the effective
string action. All these terms sum up to give the temperature dependence of the string tension and as a consequence the L\"uscher term
cannot be resolved anymore. On the contrary, all higher-order terms of the effective string action do not contribute to the logarithmic
correction, which thus maintains its free bosonic value $k=1/2$, irrespective of the details of the effective string action, and  thus it can be used as a
universal signature of the effective string description.

As mentioned above, the precision of our data is  
sufficient to reveal clear quantitative deviations from the predictions of the Nambu-Goto string. These are most clearly 
exhibited by the coefficients of the 
$\mathcal{O}(r^{-2})$ terms in the fitted expressions for $Q$, eqs.~(\ref{SU2_Q_fit}), (\ref{SU3_Q_fit}), (\ref{SU4_Q_fit}), 
(\ref{Z2_cold_Q_fit}) and 
(\ref{Z2_hot_Q_fit}), which agree in sign with the expectations from eq.~(\ref{NG_prediction_for_Q}), but are different
 in amplitude. Our data analysis shows 
that such deviations are not compatible with statistical fluctuations: rather, they reveal a real physical effect, 
namely that in this temperature range the 
idealization of the confining flux tube as a Nambu-Goto string already starts to fail.

This is not surprising and is consistent with similar observations already pointed out in the literature (for an updated review see, for instance, ref.~\cite{Teper:2009uf}).
In particular, ref.~\cite{Bialas:2009pt} also found that, for $\SU(3)$ Yang-Mills theory in $D=2+1$ dimensions, sizeable deviations from the Nambu-Goto model 
show up for temperatures higher than $0.7 T_c$---although the ``temperature-dependent string tension'' remains close to its Nambu-Goto expectation in that regime. 
Our results are in full agreement with these findings, and generalize them to groups other than $\SU(3)$.

Concerning the gauge group dependence, one interesting observation is that our results for the $\Z_2$ gauge model show somewhat more pronounced deviations w.r.t. 
the Nambu-Goto model. In particular, the constant term in eq.~(\ref{Z2_hot_Q_fit}) tends to deviate from the corresponding predictions 
for the temperature-dependent string tension $a^2 \sigma$. This agrees with previous observations~\cite{Caselle:2002ah,truncated}, that a \emph{truncated} Nambu-Goto 
prediction is actually (accidentally) in better agreement with the numerical data than the full expectation. On the other hand, the analogous results for 
the $\SU(2)$, $\SU(3)$ and $\SU(4)$ gauge groups are in better agreement with the Nambu-Goto prediction, and do not reveal a particular dependence on the 
number of colors. 

On the theoretical level, another observation is that, if one squeezed the lattice size in the direction orthogonal to the plane defined by the two Polyakov loops
(i.e. the direction in which the effective string fluctuates), then at some point, when the lattice size becomes of the same order of the effective flux tube width, 
the string would start to wind around the compactified transverse direction (we assume periodic boundary conditions in all the lattice directions). 
At this point it would be more natural to model the flux tube 
fluctuations in terms of a \emph{compactified} bosonic string theory. In this case, the leading term in the large-distance behavior of $A$ would vanish; 
this effect could be used for precision numerical tests of the flux tube width.

To summarize, in this work we have presented the results of a high-precision numerical investigation of confining gauge theories in $D=2+1$ dimensions, 
at finite temperature. We studied the behavior of the force between static sources in the fundamental representation, and compared the results of our 
lattice simulations for $\SU(2)$, $\SU(3)$ and $\SU(4)$ Yang-Mills theories, as well as for the $\Z_2$ gauge model, with the predictions of an effective 
theory based on a confining bosonic string described by the Nambu-Goto action. In particular, we focused on an observable which can be interpreted as a 
finite-temperature analogue of the L\"uscher term, and which can be obtained to high precision from correlators of Polyakov lines---provided one 
\emph{does not} perform a projection to their zero transverse-momentum component. This poses a technical difficulty, which we overcame by using 
advanced simulation algorithms, including multi-level error-reduction methods~\cite{Luscher:2001up}.

We found good agreement between the string theory model and lattice results, and a very mild dependence on the gauge group rank in $\SU(N)$ Yang-Mills theories. 
This confirms the strong similarities between these theories, which has already been remarked in other works, both in $D=2+1$ and in 
$D=3+1$. We also found clear quantitative deviations from the Nambu-Goto predictions, which show up in the subleading correction terms in the 
interquark force and its derivative. Remarkably enough, even in the cases in which we observe deviations from the Nambu-Goto expectation,
 the coefficient of the logarithm term in the potential keeps the expected value $k=1/2$ and can be used as a precise 
 signature of the effective string in a finite temperature geometry.

\vskip1.0cm {\bf Acknowledgements.}
We warmly thank  F.~Gliozzi for many useful discussions and suggestions. M.C. thanks all the participants of the {\it Confining Flux Tubes and Strings} 
Workshop at ECT$^\star$, Trento during July 2010 for several useful discussions, which partially stimulated the present work. M.P. acknowledges financial support from the Academy of Finland, project 1134018. This research was supported in
part by the European Community - Research Infrastructure Action under the
FP7 ``Capacities'' Specific Programme, project ``HadronPhysics2''.

\end{document}